\documentclass[lettersize,journal]{IEEEtran}
\usepackage{amsmath,amsfonts} % amssymb is also commonly used with these

\usepackage{array}
\usepackage[caption=false,font=normalsize,labelfont=sf,textfont=sf]{subfig}
\usepackage{textcomp}
\usepackage{stfloats}
\usepackage{algorithm}% Only once

\usepackage{url}
\usepackage{verbatim}
\usepackage{graphicx}
\usepackage{cite}
\usepackage{enumitem} % No need to load amsmath again, it's already included
\usepackage{multirow} 
\usepackage{booktabs}
\usepackage[utf8]{inputenc}
\usepackage{algpseudocode}% If you're using algpseudocode, typically you don't need algorithmic
\usepackage[
pdfauthor={derajan},
pdftitle={How to do this},
pdfstartview=XYZ,
bookmarks=true,
colorlinks=true,
linkcolor=red,
urlcolor=green,
citecolor=green,
pdftex,
bookmarks=true,
linktocpage=true,   % makes the page number as hyperlink in table of content
hyperindex=true
]{hyperref}
\usepackage{orcidlink}

\begin{document}

\title{Enhancing Collaborative Semantics of Language Model-Driven Recommendations via Graph-Aware Learning}

\author{Zhong Guan, Likang Wu~\orcidlink{0000-0002-4929-8587}, Hongke Zhao~\orcidlink{0000-0003-3099-4803}, Ming He~\orcidlink{0000-0002-1870-7472}, Jianpin Fan~\orcidlink{0000-0003-2290-1785
}
        % <-this % stops a space
\thanks{Zhong Guan, Likang Wu, Hongke Zhao are with the College of Management and Economics, Laboratory of Computation and Analytics of Complex Management Systems (CACMS),Tianjin University, Tianjin 30072, China (e-mail: gz851508778@tju.edu.cn; wlkktmzgl@gmail.com; hongke@tju.edu.cn,).}% <-this % stops a space
\thanks{Ming He and Jianping Fan are with the AI Lab at Lenovo Research, Beijing 100000, China (e-mail: heming01@foxmail.com; jfan1@lenovo.com).}
}

% The paper headers
\markboth{Journal of \LaTeX\ Class Files,~Vol.~14, No.~8, August~2021}%
{Shell \MakeLowercase{\textit{et al.}}: A Sample Article Using IEEEtran.cls for IEEE Journals}

%\IEEEpubid{0000--0000/00\$00.00~\copyright~2021 IEEE}
% Remember, if you use this you must call \IEEEpubidadjcol in the second
% column for its text to clear the IEEEpubid mark.

\maketitle

\begin{abstract}
Large Language Models (LLMs) are increasingly prominent in the recommendation systems domain. Existing studies usually utilize in-context learning or supervised fine-tuning on task-specific data to align LLMs into recommendations. However, the substantial bias in semantic spaces between language processing tasks and recommendation tasks poses a nonnegligible challenge. Specifically, without the adequate capturing ability of collaborative information, existing modeling paradigms struggle to capture behavior patterns within community groups, leading to LLMs' ineffectiveness in discerning implicit interaction semantic in recommendation scenarios. To address this, we consider enhancing the learning capability of language model-driven recommendation models for structured data, specifically by utilizing interaction graphs rich in collaborative semantics. We propose a \textbf{G}raph-\textbf{A}ware Learning for \textbf{L}anguage Model-Driven \textbf{Rec}ommendations (GAL-Rec). GAL-Rec enhances the understanding of user-item collaborative semantics by imitating the intent of Graph Neural Networks (GNNs) to aggregate multi-hop information, thereby fully exploiting the substantial learning capacity of LLMs to independently address the complex graphs in the recommendation system. Sufficient experimental results on three real-world datasets demonstrate that GAL-Rec significantly enhances the comprehension of collaborative semantics, and improves recommendation performance.
\end{abstract}


\section{Introduction}
\IEEEPARstart{L}{arge} Language Models (LLMs) like ChatGPT \cite{26,31} and LLaMA~\cite{32} have shown remarkable proficiency in tasks involving semantic understanding and knowledge inference. Consequently, LLMs are now being employed in a diverse range of tasks and domains ~\cite{23,24,14,9,a,b,c}. Recommendation systems are anticipated to derive substantial advantages from the advancements in LLMs.

Motivated by these developments, employing LLMs is increasingly becoming a focal point in recommendation system research. This involves positioning LLMs as predictors in recommendation systems through prompts based on natural language~\cite{3,25,d}. Initially, LLMs were engaged in recommendation tasks via in-context learning~\cite{26,4}. However, substantial research has shown that LLMs, when pre-trained on general natural language corpora, lack the domain-specific knowledge and capabilities required for recommendation tasks~\cite{1,3}. Consequently, there is a shift towards fine-tuning LLMs using domain-specific recommendation data to significantly enhance their performance~\cite{6,7,8}. Yet, these methods of fine-tuning can't bridge the performance gap between LLMs and traditional recommendation models. LLMs, trained within a natural language paradigm, primarily focus on acquiring explicit knowledge but still fall short in understanding implicit interactions within recommendation systems, thereby failing to mine the interaction relationship between users and items.
\begin{figure}[t]
  \centering
  \includegraphics[width=\linewidth]{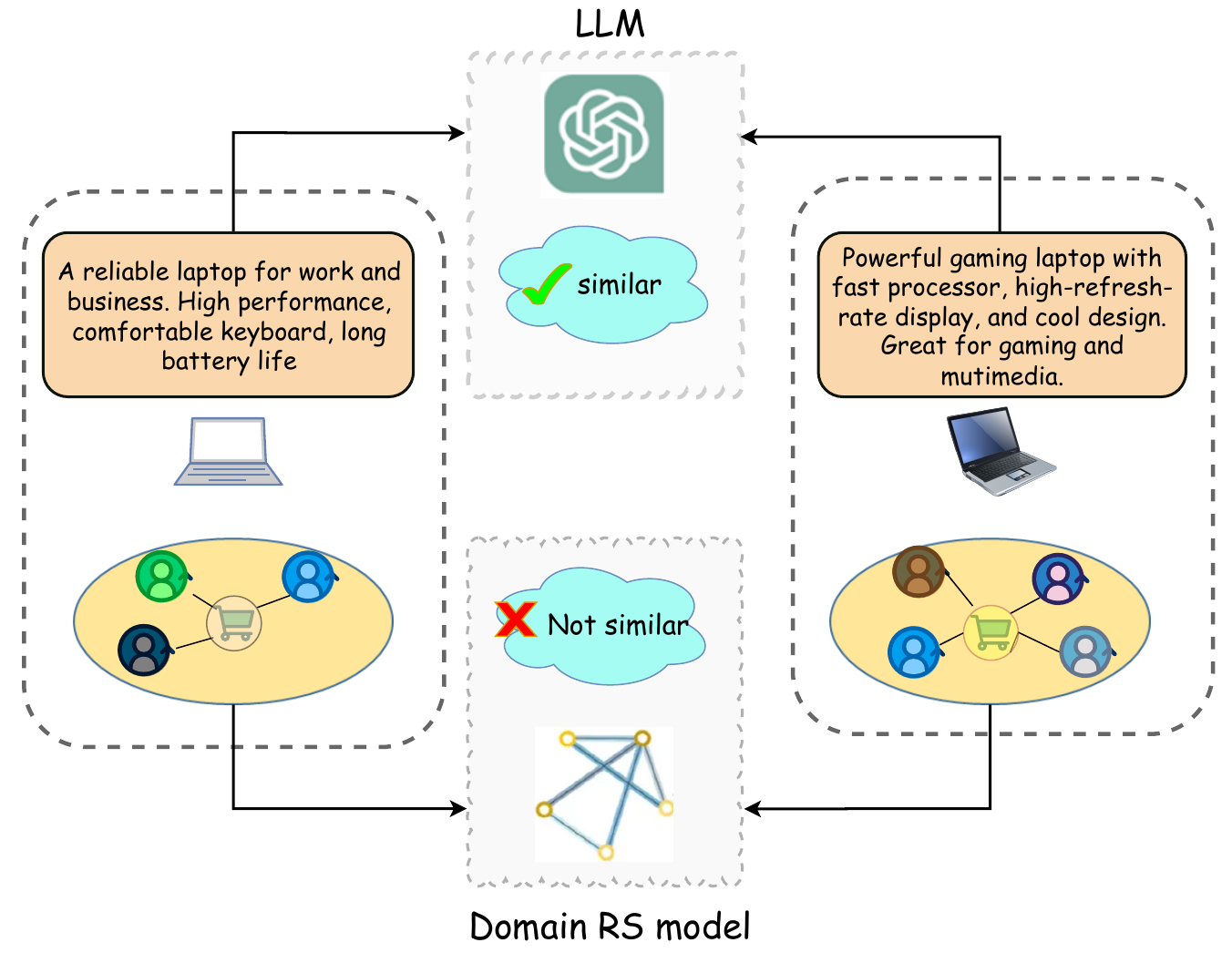}
  \caption{The differences between LLMs and traditional models in distinguishing items.}
  \label{fig:figure1}
\end{figure}

Within recommendation scenarios, the interaction data between users and items manifest as interaction graphs. Current LLMs are not adept at extracting implicit information from these interaction graph datasets. Utilizing LLMs in recommendation systems encounters two significant issues:

\noindent\textit{1). A misalignment exists between the semantic space of LLMs and the semantic space of user-item collaborative information.}

The semantic space of natural language is inconsistent with the semantic space of collaborative information~\cite{48,7}, and merely through supervised fine-tuning, it is difficult to align different semantic spaces. This alignment challenge hinders LLMs from identifying a unified semantic subspace. Consequently, natural language-based LLMs struggle to distinguish similar items with subtle differences. As shown in Fig. \ref{fig:figure1}, in the same broad category, items with minor textual differences accessed by different users show clear distinctions in the semantic space of collaborative information, but appear more consistent in the semantic space of LLMs. 

\noindent\textit{2). The training paradigms of LLMs are not well-suited to align with the behavioral patterns observed in recommendation system communities.}

While LLMs are adept at learning frequent patterns that can be formalized as knowledge, they are less effective in mining and leveraging the behavioral patterns prevalent within communities~\cite{16,8}. Specifically, LLMs face challenges in modeling implicit user-item interaction information. In contrast, representative models like GNNs are capable of effectively modeling the topological structures within user-item bipartite graphs.

Therefore, to enable LLMs to capture semantic information in graph structures, we introduce the GAL-Rec framework, which jointly models the language understanding capabilities of LLMs with the aggregation intent of GNNs in the field of graph modeling. However, using a general strategy of simply and directly fusing the representations of the two does not allow the language model to learn structural modal knowledge and therefore cannot align the representation space. When designing learning paradigms, we have encountered the following technical challenges: 

i). How can LLMs be enabled to recognize the complex structure of user-item graphs?

ii). How can the issue of negative sample collection for LLMs be addressed under graph-aware learning?

Initially, to enhance LLMs' receptive field and capture multi-hop information, we devised a prompt structure that ingeniously includes both users' past interactions and data from users sharing common items, mirroring the incorporation of one- and two-hop neighbors in a graph. Subsequently, to enhance LLM's comprehension of collaborative information, we were inspired by the aggregation strategy employed in GNNs, adopting a graph-aware contrastive learning approach. This strategy enables LLMs to explicitly assimilate the interaction dynamics between users and items, thus improving LLM's performance through the implementation of GNN's aggregation techniques on bipartite graphs. And  it suggests that a user's two-hop information (\textit{user-2-hop}) should align closely with their own profile (\textit{user-0-hop}). Similarly, a user's one-hop information \textit{(user-1-hop) }should align with the information of the item they interacted with (\textit{item-0-hop}), and the user's two-hop information (\textit{user-2-hop}) should align with the one-hop information of the item (\textit{item-1-hop}). The same holds true for items. After constructing positive sample pairs through the aforementioned operations, we employ dynamic queue storage based on Moco~\cite{46} to collect and update the set of negative samples, catering to the requirements of graph-aware learning, a contrastive learning strategy.
This framework, which draws inspiration from GNN's aggregation methodology and graph contrastive learning~\cite{he2023candidate}, facilitates a deeper understanding of collaborative embeddings in LLMs.

In summary, we conclude our main contributions as follows:
\begin{itemize}[leftmargin=*]
\item We introduced a novel (GAL-Rec) framework, designed to facilitate graph-aware learning for enhanced comprehension of collaborative information by LLMs, thereby enabling LLMRec to understand the interaction graph data in recommendation system.
\item We first proposed graph-aware learning  for collaborative semantics extraction in GAL-Rec, fully harnessing the vast potential of LLMs and aligning them with recommendations in a self-supervised manner.
\item Extensive experimental results on real-world datasets demonstrate that our proposed GAL outperforms several state-of-the-art models in terms of performance.
\end{itemize}

\section{Related Works}
In this section, we review the applications of LLMs in recommendation systems as well as the work that combines LLMs with GNNs.
\subsection{LLM-based Recommendations}
Recently, there has been a growing focus on integrating large language models into recommendation systems, owing to the emergence of these powerful models~\cite{3,7}. Here we introduce some representative works. Initially, researchers used public APIs like ChatGPT for in-context learning and implemented recommendation system applications through conversational interaction functions~\cite{3}. However, due to the significant gap between recommendation tasks and LLM's pre-training, in-context learning could only have good effects in cold-start scenarios and was greatly limited in warm-start scenarios~\cite{4}. To enhance LLMs' capabilities in recommendation contexts, some studies have focused on fine-tuning them with recommendation-specific datasets~\cite{1,3,wu2023survey}. For instance, P5~\cite{1} relies on the tokenizer used by LLM (Sentence Piece tokenizer) to generate item ID tokens, M6~\cite{5} considers using item text information (names) as item IDs, Shashank~\cite{6} explores the use of semantic ID models for generative recommendation systems and proves their effectiveness, applying special quantization methods to the textual embeddings of items to form semantic IDs for items.

However, simple directive-based fine-tuning is insufficient to bridge the substantial gap between recommendation tasks and the inherent capabilities of LLMs. LLMs still lack an understanding of collaborative information and insufficient learning of implicit interaction information in recommendation tasks. CoLLM~\cite{7} adopts a two-tiered fine-tuning strategy for LLMs. The first stage is aimed at acclimatizing the LLMs to the specificities of recommendation tasks. The second stage is focused on imparting an understanding of collaborative information. Contrastively, Qiu~\cite{8} employs contrastive learning across diverse instructional datasets to more efficiently align the representations of natural language and collaborative information in LLMs. Huang~\cite{Hua_2023} et al. investigated various methods of injecting collaborative signals into LLMs by encoding item IDs. Different from the above works, our architecture (GAL-Rec) encourages self-supervised learning and understanding of collaborative information through a contrastive learning framework, enabling LLMs to effectively utilize and further improve the collaborative information captured by traditional recommendation models.

\subsection{Integration of GNNs with LLMs}
The integration of LLMs with GNNs in existing research primarily falls into two categories:

i). LLM as Enhancer.
 Initially, LLMs were primarily employed for feature encoding, where their embeddings were utilized as node features in graphs for GNN training~\cite{10,11,12}. Subsequent developments led to the joint training and collaborative learning of LLMs and GNNs~\cite{13,9,14,15}. For instance, GLEM~\cite{9} integrates LLMs and GNNs using a pseudo-likelihood variational framework, allowing each component to be trained independently, thus ensuring scalability and effectiveness. OFA~\cite{18} unifies various datasets and tasks by leveraging LLMs to represent node information, introducing node prompts and graph prompt paradigms for cross-task and dataset learning in GNNs. TAPE~\cite{he2023harnessing} leverages advanced LLMs for textual analysis of Text-Attribute Graphs, with the derived embeddings subsequently utilized for a variety of tasks in GNNs. LLMRG~\cite{19} employs large language models for graph structure enhancement, specifically augmenting user-item interaction edges, item node attributes, and user profiles. This approach explicitly infers user-item interaction patterns, addressing the challenge of sparse implicit feedback signals.
 ii). LLM as the Predictor.
 The success of LLMs has also sparked interest in using them as predictors  for addressing graph-structured tasks. Research like~\cite{16,17,51,52,wu2023learning}focuses on describing graph structures through textual language, aiming for LLMs to capture graph-related signals. GraphText~\cite{20} introduces a versatile graph reasoning framework for context learning and directive fine-tuning of graph structures. GLRec~\cite{21} creates a meta-path prompt constructor, facilitating LLMs' understanding of varied behavioral information sequences in heterogeneous graphs. Despite these advancements, LLMs struggle with structured data translated into natural language, often yielding suboptimal results. To overcome this challenge, LLaGA~\cite{chen2024llaga} reformulates node link information as sequential data, applying instruction tuning that LLMs can comprehend while maintaining structural node information. TP~\cite{22} simulated the message propagation process of GNN, utilized LLM to propose and aggregate similar questions, promoting the form of "message propagation" in response generation. In contrast, our method places greater emphasis on enabling LLMs to actively learn the GNN aggregation paradigm on bipartite graphs, thereby augmenting LLM performance. Furthermore, the above studies predominantly explore LLMs as predictors within the context of text-graph scenarios, focusing on node classification as the predictive task, our investigation delves deeper into the application of LLMs as predictors within recommendation systems.

\section{PRELIMINARY}

\begin{figure*}[t]
\centering
\includegraphics[width=0.9\textwidth]{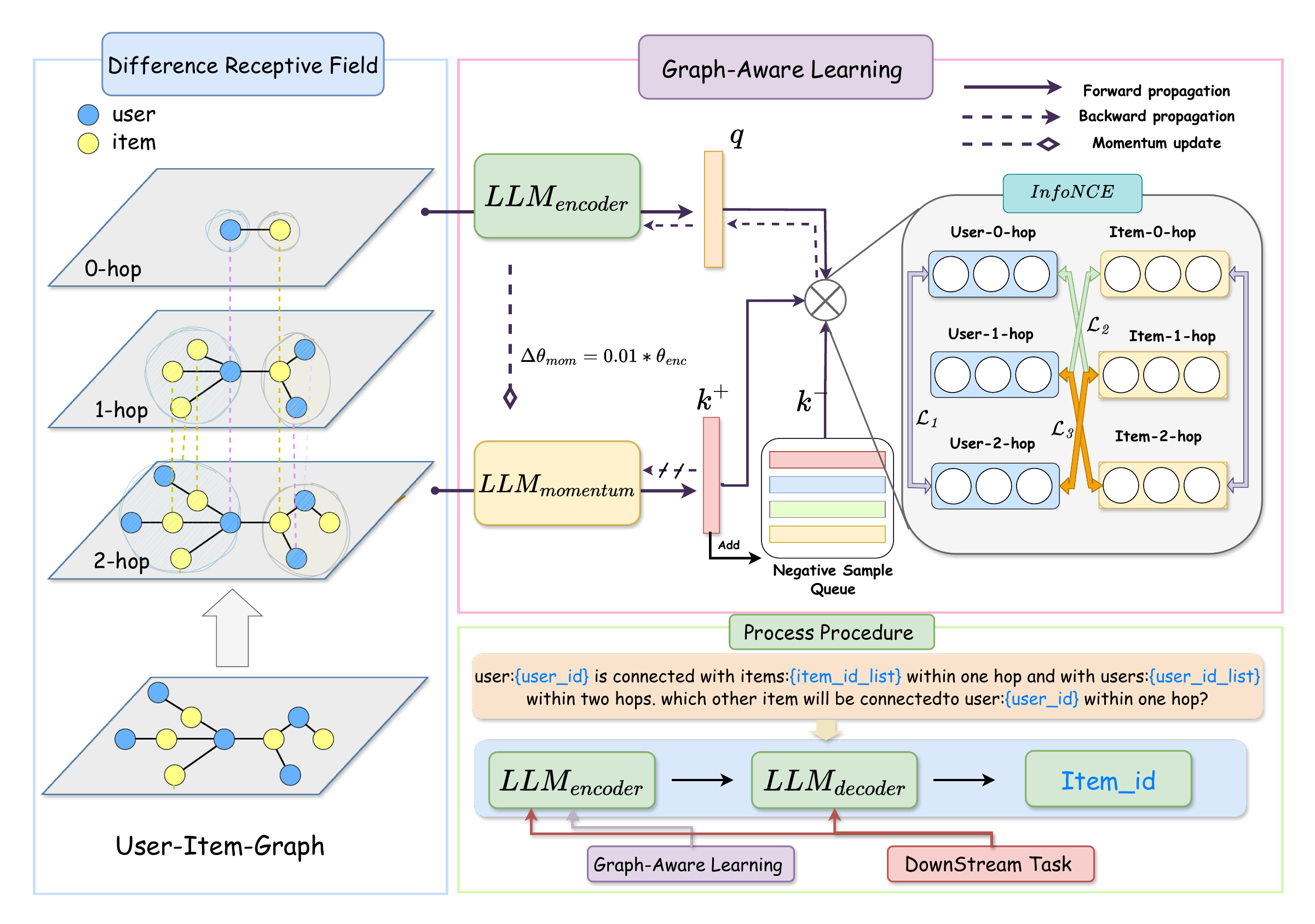}
\caption{An overall illustration of our proposed learning architecture. The left panel shows the selection of information with different hops, and the right performs graph-aware learning.}
\label{fig:figure2}
\end{figure*}
\subsection{Problem Definition}
Consider two set of: $\mathcal{U}$=\{$u_1$, $u_2$, …, $u_n$\} and $\mathcal{I}$=\{$i_1$, $i_2$, …, $i_m$\}, representing $N$ users and $M$ items, respectively. Generally, a user $u$ has a chronologically-ordered interaction sequence with items: \{$i_1$,...,$i_k$\}, where $k$ is the number of interactions and $i_t$ is the t-th item that the user $u$ has interacted with. Based on above notations, we can define the task of sequential recommendation. Formally, given the historical behavior of a user \{$i_1$,...,$i_k$\}, the sequential recommendation task is to predict the next item that the user is likely to interact with at the $(k+1)$-th step. 

Our objective is to enhance the recommendation performance by enabling large language models to better understand collaborative information.
\subsection{Instruction Tuning}
In the process of instruction tuning, we initially assign unique index identifiers to each item $i$$\in$ $\mathcal{I}$ and user $u$$\in$ $\mathcal{U}$ within the recommendation dataset. These identifiers are derived by augmenting the existing dataset.Subsequently, we construct instruction data pairs $(x, y)$. Here, $x$ constitutes the instructional input, formulated from the multi-layered relational information derived from the user’s interaction sequence with items, and $y$ represents the instructional output. Lastly, the optimization of the LLM is conducted. For the provided data pairs $(x, y)$, the LLM employs the negative log-likelihood method for its optimization process:
\begin{equation}
\label{eq:eq1}
\mathcal{L}_{0} = -{\sum_{t=1}^{|y|}}log P_\theta(y_t|y_{<t},x),
\end{equation}
where $x$ and $y$ represent the Instruction Input and Instruction Output, respectively. And $y_t$ is the t-th token of the $y$, $y_{<t}$ represents the tokens before $y_t$, $\theta$ is the parameters of LLM.

\section{METHODOLOGY}
In our study, we aim to enhance LLM's proficiency in collaborative information understanding and improve their efficacy in recommendation tasks. This goal is pursued by learning from GNN's aggregation patterns and understanding the dynamics of bipartite graphs in recommendation contexts.

The model we propose is shown in Fig. \ref{fig:figure2}, which comprises four primary components: (1). External Embeddings: Introduce traditional model embeddings or text embeddings to facilitate the initialization of new identifiers within the expansion of the LLM corpus. (2). Prompt Construction: Inspired by the aggregation of GNNs, we have developed a prompt structure that incorporates multi-hop information. (3). Graph-aware learning(GAL) module for LLMs: by leveraging the Graph-aware learning module, we enabled the LLM to effectively model multi-hop information, thereby significantly enhancing its capacity to comprehend collaborative information. (4). Dynamic Queue Storage: to address the challenge of negative sample collection posed by the large scale of LLMs, we employ a dynamic queue storage strategy based on Moco.

By integrating these four components, our goal is to enhance LLM's understanding of collaborative information, thereby improving its performance in recommendation tasks.
\subsection{External Embeddings}

To facilitate the model's handling of recommendation task inputs, the corpus is meticulously designed to include specific tokens for each user and item. We uniquely identify users and items by extending the corpus to incorporate new tokens representing them within the corpus. In the process of embedding new tokens, we utilize two distinct methodologies: semantic information embeddings derived from item texts, and collaborative information embeddings obtained from traditional recommendation models.

\begin{itemize}[leftmargin=*]
\item \emph{Text Embeddings}. To provide LLMs with semantic information about items, we assume that each item has associated content features that can capture useful semantic information (such as titles, descriptions, or brands). These features are initially concatenated into a comprehensive prompt, which is subsequently processed by a pre-trained text encoder (in this case, Sentence-T5). This process transforms the textual information of items into semantic embeddings, suitable for integration into LLM. Formally, we have:
\begin{equation}
\begin{aligned}
&x_i = \text{Concat}(x^{Title}_i, x^{Des}_i, x^{Cate}_i, x^{Brand}_i) \\
&e_i = \text{Sentence}_{T5}(x_i).
\end{aligned}
\end{equation}
\item \emph{Traditional Model Embeddings}. To further enrich large models with user-item collaborative information, we leverage traditional recommendation models to extract such information. User-item representations, as captured by these models, are then mapped into the token embedding space of LLMs, and instance representations can be obtained as follows:
\begin{equation}
\begin{aligned}
\label{eq:eq3}
e_u &= g_{\theta}(h_u), &\quad h_u &= f_{\phi}(u),
\\
e_i &= g_{\theta}(h_i), &\quad h_i &= f_{\phi}(i),
\end{aligned}
\end{equation}
where $f_{\phi}$ represents the traditional recommendation model, while $h_u$ denotes the user embeddings derived from this model. Furthermore, a mapping layer, denoted as $g_{\theta}$, is employed to convert $h_u$ , which has a dimension of $d_1$ , into e with a dimension of $d_2$. This transformation aligns the dimensions of e with those of the embeddings in the LLMs, where typically $d_1$ is less than $d_2$. And the same applies to item $i$.
    
\end{itemize}

\subsection{Prompt Construction}

Similar to existing methods~\cite{1}, GAL also constructs prompt language based on recommendation data, and LLM generates recommendations based on this prompt language. The key distinction in our method lies in our adaptation of the prompt construction to more closely resemble the aggregation processes observed in GNNs.

\begin{figure}[h]
  \centering
  \includegraphics[width=\linewidth]{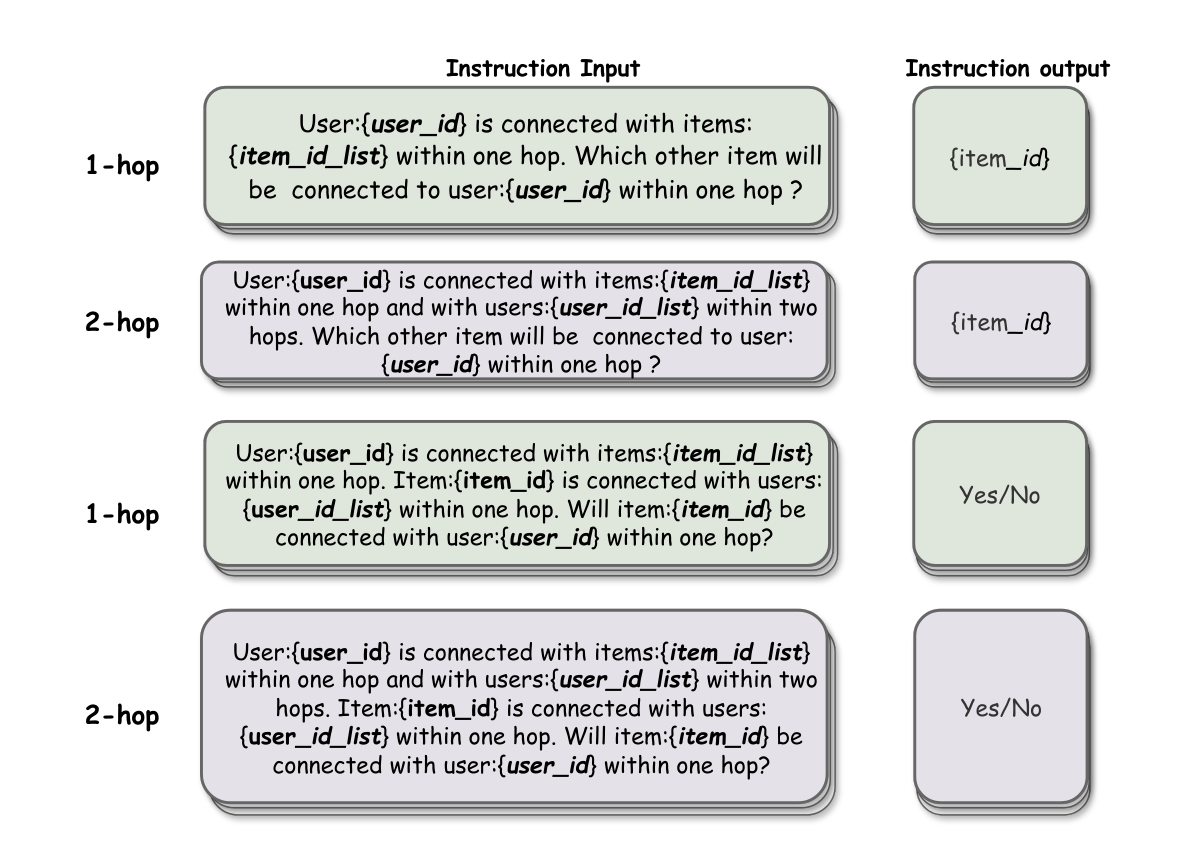}
  \caption{ A case demonstrating the motivation of designing task-specific tokens.}
  \label{fig:figure3}
\end{figure}

Echoing the GNN strategy of aggregating information from multi-hop neighbors, our recommendation process for a user involves providing the LLM not only with the user's sequence of item interactions (1-hop information) but also with data about other users who have interacted with the same items (2-hop information). This approach is designed to enhance the LLM's capacity to generate more contextually rich and relevant recommendations by incorporating a broader spectrum of user-item interaction data. The structure of the prompt template is depicted in Fig. \ref{fig:figure3}.

As an illustration, consider the 2-hop scenario in the template: $user\_id$ serves as the user's unique identifier, $item\_id\_list$ denotes the array of items with which the user has interacted, and $user\_id\_list$ signifies the collection of users who have engaged with the same items as the given $user\_id$, effectively representing second-order nodes in the interaction graph. Within this framework, $ID\_Embedding$ is utilized to incorporate collaborative information as captured by the conventional model or the semantic information processed by a pre-trained model.
Then we further encode the instruction data into the embedding sequence \textit{E:}

\begin{equation}
\label{eq:eq4}
E=[e_1,e_2,...,e_u,..,e_i,...],
\end{equation}

where  \(e_u\) is obtained from $e_1$=$Emb_{LLM}(x_1)$. While $e_u/e_i$ denotes the external embeddings
\subsection{Graph-Aware Learning}

In this section, we will introduce GAL. As previously discussed, merely mapping collaborative information captured by external models into an LLM’s token embedding space for fine-tuning proves insufficient. This approach does not enable  LLMs to fully grasp collaborative signals, leading to suboptimal performance. Furthermore, solely embedding items' semantic information into LLMs, while neglecting the incorporation of collaborative information, overlooks critical aspects of user-item interactions. Consequently, we advocate for a contrastive learning methodology for LLM, drawing inspiration from GNN's aggregation strategies.

In the user interaction graph, two distinct node types exist: users and items. Invariably, a user's first-order neighbors are items, and their second-order neighbors are other users, a property mirrored in item nodes.

In the context of collaborative filtering within recommendation systems, we observe the following on the user side:

\noindent\textit{1). Aligning user-2-hop with user-0-hop. }
In a user-item interaction graph, second-order neighbors of a central user (item) are defined as those similar, like-minded users (items) who share common features or preferences. For the central node, the process of aggregating second-order information essentially entails consolidating the features of these similar homogeneous neighbors. Within the recommendation system's semantic space, this aggregated data is expected to closely resemble, and align more with, the central user's or item's embedding. Therefore, in the context of integrating LLMs within recommendation systems, the formal expression is delineated as follows:

\begin{equation}
\label{eq:eq5}
\small
\mathcal{L}^u_1 = -\log\frac{\exp\Big(\text{sim}\Big(E^0_u, E^2_u\Big)/\tau\Big)}{\sum_{v \in \mathcal{U}}\exp\Big(\text{sim}\Big(E^0_v, E^2_u\Big)/\tau\Big)},
\end{equation}

\begin{equation}
\label{eq:eq6}
\small
\mathcal{L}^i_1 = -\log\frac{\exp\left(\text{sim}\left(E^0_i, E^2_i\right)/\tau\right)}{\sum_{j \in \mathcal{I}}\exp\left(\text{sim}\left(E^0_j, E^2_i\right)/\tau\right)},
\end{equation}

\begin{equation}
\label{eq:eq7}
\small
\mathcal{L}_1 = \mathcal{L}^u_1+\mathcal{L}^i_1,
\end{equation}
wherein $E^0_u$ is the embedding of the user's zero-order aggregated data processed by the LLM-encoder,  $E^2_u$ is the embedding of the user's second-order aggregated data processed by the LLM-encoder, $E^0_v$ is the embedding of the other user's zero-order aggregated data processed by the LLM-encoder ,$\tau$ is the temperature hyperparameter of softmax. For item i, $E^0_i$/$E^2_i$ are similar.

\begin{algorithm}[t]
\caption{Algorithm of Graph-aware learning}
\begin{algorithmic}[1]
\renewcommand{\algorithmicrequire}{\textbf{Input:}}
\renewcommand{\algorithmicensure}{\textbf{Output:}}
\Require user set $U$, item set $I$, interactions set $R$,prompt $P$,external embedding $h_u$, $h_i$
\Ensure LLM parameters $\theta_{llm}$, mapping layer parameters $\theta_{g}$
\State Extend user/item tokens for LLM
\State Initialize mapping layer $\theta_{g}$, $\theta_{llm}$ is initialized with pretrained LLM parameters
\While{ not converge}
  \State Sample($u_0$,$i_0$) $\in$ R, prompt P, external embedding $h_u$, $h_i$
  \State Sample multi-hop information $u_1$,$u_2$,$i_1$,$i_2$ from R
  \State // Forward propagation
  \State Obtain $e_u$, $e_i$ via Eq.(\ref{eq:eq3})
  \State Obtain Embedding $E^0_u$, $E^1_u$, $E^2_u$, $E^0_i$, $E^1_i$, $E^2_i$ using LLM-Encoder
  \State // Calculate Loss
  \State Calculate  $\mathcal{L}_1$ via Eq.(\ref{eq:eq5}) (\ref{eq:eq6}) 
    (\ref{eq:eq7})
  \State Calculate  $\mathcal{L}_2$ via Eq.(\ref{eq:eq8}) (\ref{eq:eq9}) 
   (\ref{eq:eq10})
  \State Calculate  $\mathcal{L}_3$ via Eq.(\ref{eq:eq11}) (\ref{eq:eq12}) (\ref{eq:eq13})
  \State Calculate Cross Entropy loss $\mathcal{L}_0$ via Eq.(\ref{eq:eq1})
  \State Calculate finial loss $\mathcal{L}_{GAL}$ via Eq.(\ref{eq:eq13})
  \State // Back propagation
  \State Update $\Theta_{llm} \gets \Theta_{llm} - \alpha \cdot \frac{\partial L_{GAL}}{\partial \Theta_{llm}}$
  \State Update $\Theta_{g} \gets \Theta_{g} - \alpha \cdot \frac{\partial L_{GAL}}{\partial \Theta_{g}}$
\EndWhile
\State \textbf{return} $\theta_{g}$ ,$\theta_{llm}$
\end{algorithmic}
\end{algorithm}

\noindent\textit{2). Aligning user-2-hop with item-1-hop. } In this scenario, both the second-order neighbors of a user and the first-order neighbors of adjacent items are categorized as homogeneous nodes, implying they share similar attributes or categories. During the aggregation process in GNN, these nodes are interconnected during aggregation, forming a dependency structure that spans over greater distances. Specifically, the second-order aggregated information for a user summarizes the features of these similar users, while the first-order aggregation of the neighboring items indirectly mirrors the preferences of like-minded users. Hence, at the semantic level of a recommendation system, these aggregated representations should exhibit high similarity and consistency. Specifically, within the encoder module of a LLM, this is manifested as:

\begin{equation}
\label{eq:eq8}
\small
\mathcal{L}^u_2 = -\log\frac{\exp\Big(\text{sim}\Big(E^2_u, E^1_i\Big)/\tau\Big)}{\sum_{v \in \mathcal{U}}\exp\Big(\text{sim}\Big(E^2_v, E^1_i\Big)/\tau\Big)},
\end{equation}
\begin{equation}
\label{eq:eq9}
\small
\mathcal{L}^i_2 = -\log\frac{\exp\Big(\text{sim}\Big(E^2_i, E^1_u\Big)/\tau\Big)}{\sum_{j \in \mathcal{I}}\exp\left(\text{sim}\left(E^2_j, E^1_u\right)/\tau\right)},
\end{equation}
\begin{equation}
\label{eq:eq10}
\small
\mathcal{L}_2 = \mathcal{L}^u_2+\mathcal{L}^i_2,
\end{equation}
wherein $E^2_u$ is the embedding of the user's second-order aggregated data processed by the LLM-encoder, $E^1_i$ is the embedding of the item's first-order aggregated data processed by the LLM-encoder, $E^2_v$ is the embedding of the other item's first-order aggregated data processed by the LLM-encoder, $E^2_i$/$E^1_u$/$E^2_j$ are similar.

\noindent\textit{3). Aligning user-1-hop with item-0-hop. } Reflecting the foundational hypothesis of recommendation systems, items sharing similarities are more inclined to engage with the same user groups. Thus, a particular neighboring item of a user is likely to exhibit a general similarity with the user’s historically interacted items. Accordingly, the first-order aggregate information of a user should bear a resemblance to the intrinsic data of a specific neighboring item. In formal terms, this concept is articulated as:
\begin{equation}
\label{eq:eq11}
\small
\mathcal{L}^u_3 = -\log\frac{\exp\Big(\text{sim}\Big(E^0_u, E^1_i\Big)/\tau\Big)}{\sum_{v \in \mathcal{U}}\exp\Big(\text{sim}\Big(E^0_v, E^1_i\Big)/\tau\Big)},
\end{equation}
\begin{equation}
\label{eq:eq12}
\small
\mathcal{L}^i_3 = -\log\frac{\exp\Big(\text{sim}\Big(E^0_i, E^1_u\Big)/\tau\Big)}{\sum_{j \in \mathcal{I}}\exp\Big(\text{sim}\Big(E^0_j, E^1_u\Big)/\tau\Big)},
\end{equation}
\begin{equation}
\label{eq:eq13}
\small
\mathcal{L}_3 = \mathcal{L}^u_3+\mathcal{L}^i_3,
\end{equation}
wherein $E^0_u$ is the embedding of the user's second-order aggregated data processed by the LLM-encoder, $E^1_i$ is the embedding of the item's first-order aggregated data processed by the LLM-encoder, $E^0_v$ is the embedding of the other user's first-order aggregated data processed by the LLM-encoder, $E^0_i$/$E^1_u$/$E^0_j$ are similar.

\subsection{MoCo-based Dynamic Queue Storage}
In graph-aware learning, we adopt contrastive learning to empower LLM to capture the relationships among multi-hop information in complex graph structures. However, the efficacy of contrastive learning is highly contingent upon the quantity of negative samples, posing a unique challenge for LLM applications. The standard tactic of inflating batch sizes to augment the negative sample count encounters limitations when confronted with the colossal parameter dimensions and substantial GPU memory demands characteristic of LLMs. To address this issue, we employ a dynamically queued storage mechanism inspired by MoCo.

This ingeniously bypasses the limitations imposed by batch size, operating through two pivotal mechanisms: first, the establishment and maintenance of a dynamic memory queue (memory bank) for accumulating and updating the collection of negative examples; secondly, ensuring consistency in negative sample aggregation through momentum updating of the encoder. Specifically, a queue $Q$ of length $L$, a momentum encoder $f_m$, and a query encoder $f_q$ are constructed, where $f_q$ serves as the encoder component of the LLM dedicated to the primary recommendation task. Within graph-aware learning, embeddings extracted by $f_q$ are utilized as positive samples and contrasted against negative samples from queue $Q_m$. Concurrently, embeddings extracted using $f_m$ are stored into $f_m$, with the parameters of $f_m$ being updated via momentum.

Taking Eq(\ref{eq:eq5}) as an example, it is formally expressed as follows:
\begin{equation}
\label{eq:eq1000}
E_u^0=f_q(x_u^0),
E_u^2=f_q(x_u^2),
E_v^2=Sample(Q_u^2),
\end{equation}
\begin{equation}
\label{eq:eq1001}
\mathcal{L}^u_1 = -\log\frac{\exp\Big(\text{sim}\Big(E^0_u, E^2_u\Big)/\tau\Big)}{\sum_{v \in \mathcal{U}}\exp\Big(\text{sim}\Big(E^0_v, E^2_u\Big)/\tau\Big)},
\end{equation}
\begin{equation}
\label{eq:eq1002}
Q_u^2=Update(Q_u^2,f_k(x_u^2))
\end{equation}
\begin{equation}
\label{eq:eq1003}
\theta_k=m\theta_k+(1-m)\theta_q
\end{equation}
Wherein,  $\theta_k$ updates its parameters solely based on Eq(\ref{eq:eq1003}), whereas $\theta_q$ employs gradient backpropagation for parameter updates, with $m$ representing the momentum coefficient. $E_u^0$ and $E_u^2$ are positive sample pairs, $E_v^2$ is negative samples.

\subsection{Optimization}
In our approach, the three contrastive learning losses are treated as auxiliary losses. These are integrated with CE loss for joint optimization. the comprehensive loss function is formulated as follows:
\begin{equation}
\label{eq:eq14}
\mathcal{L}_{GAL} = \mathcal{L}_0+\lambda_1\mathcal{L}_1+\lambda_2\mathcal{L}_2+\lambda_3\mathcal{L}_3+\lambda_4||\Theta||^2_2
\end{equation}
where $\lambda_1$,$\lambda_2$, and $\lambda_3$ are hyperparameters to control the weights of three contrastive learning losses, respectively. Besides, $\lambda_4$ is the regularization coefficient, and $\Theta$ indicates the trainable parameters.
\linespread{1.2}
\begin{table*}[t!]
\centering
\caption{Performance comparison on sequential recommendation.}
\label{table2}
\setlength{\tabcolsep}{3pt}
\begin{tabular}{lcccccccccccc}
\toprule
\multirow{2}{*}{Methods} & \multicolumn{4}{c}{Yelp} & \multicolumn{4}{c}{Beauty} & \multicolumn{4}{c}{Toys} \\ 
\cmidrule(lr){2-5} \cmidrule(lr){6-9} \cmidrule(l){10-13}
                          & HR@5   & NDCG@5 & HR@10  & NDCG@10 & HR@5   & NDCG@5 & HR@10  & NDCG@10 & HR@5   & NDCG@5 & HR@10  & NDCG@10 \\ \midrule
Caser                    & 0.0151    & 0.0096    & 0.0253    & 0.0129     & 0.0205    & 0.0131    & 0.0347    & 0.0176     & 0.0166    & 0.0107    & 0.0270    & 0.0141     \\
BERT4Rec                      & 0.0051    & 0.0033    & 0.0090    & 0.0045     & 0.0203    & 0.0124    & 0.0347    & 0.0170     & 0.0116    & 0.0071    & 0.0203    & 0.0099     \\  
LightGCN   &  0.0255   &  0.0163   & 0.0427    &   0.0218   & 0.0305    & 0.0194    & 0.0511    & 0.0260     & 0.0270    & 0.0189    & 0.440    & 0.0234     \\  
SASRec                      &0.0162 &0.0100 &0.0274& 0.0136& 0.0387& 0.0249 &0.0605& 0.0318& 0.0463 &0.0306& 0.0675& 0.0374 \\  
S3-Rec                      &0.0201 &0.0123 & 0.0341 &0.0168 &0.0387 &0.0244 &0.0647 & 0.0327 &0.0443& 0.0294& 0.0700 &0.0376  \\  
P5                      & 0.0051& 0.0041& 0.0095& 0.0053& 0.0379& 0.025 &0.0582& 0.0298& 0.0198 &0.013& 0.0267& 0.015 \\  
CID                      & 0.0261&0.0171 &0.0428&0.0225 &0.0489 &0.0318  &0.0680&0.0357 & 0.0143 & 0.0113&0.0256 &0.0141  \\  
IID                      & 0.0232&0.0146 &0.0393& 0.0197&0.0394&0.0268  &0.0615&0.0341 & 0.131 &0.105&0.222 &0.128  \\ 
SID                      & 0.0346&0.0242 &0.0486& 0.0287&0.0430&0.0288  &0.0602&0.0368 & 0.0164 & 0.0120 & 0.0218 & 0.0139  \\ 
\hline
\footnotesize GAL(SASRec)     &0.0216 &0.0154 &0.0329  &0.0198 & 0.0486& 0.0311& 0.0759& 0.0399 &\textbf{0.0491} &\textbf{0.0322} &\textbf{0.0712}  & \textbf{0.0393} \\  
\footnotesize GAL(LightGCN)                  & 0.0323 &0.0213 & 0.0457 & 0.0256 & 0.0334 &0.0214 & 0.0537&0.0279  &0.0289&0.0190 &0.0448  &0.0241  \\  
\footnotesize GAL(Text)                      &\textbf{0.0361} &\textbf{0.0258} &\textbf{0.0491}  &\textbf{0.0298} & \textbf{0.0492}& \textbf{0.0320}& \textbf{0.0768}& \textbf{0.0406} &0.0388& 0.0260&0.0587&0.0325  \\  
\bottomrule
\end{tabular}
\end{table*}

\section{EXPERIMENTS}
In this section, we aim to address several key research questions (RQs) through comprehensive experimentation:

\begin{itemize}[leftmargin=*]
\item RQ1: Does our GAL-Rec model outperform current state-of-the-art traditional recommendation models?

\item RQ2: Is our framework capable of significantly improving the utilization of collaborative information by LLMs?

\item RQ3: Can GAL-Rec improve LLM's distribution of user (item) multi-hop information representation?

\item RQ4: How do different parameter settings affect the performance of GAL?
\end{itemize}

\subsection{Experimental Setup}
\emph{\textbf{Datasets. }} In line with the experimental approach used in P5, we will experiment on three popular benchmark datasets. These include (1). the Amazon Beauty dataset \cite{40}, comprising data from Amazon's beauty products, and (2). the Toys dataset, also derived from Amazon reviews, featuring a range of toys, and (3). the Yelp datasets. For these datasets, we will follow the methodology outlined in \cite{1}, employing the leave-one-out approach for dividing the data into training, validation, and test sets. Additionally, we filter out users and items with less than five interactions. The specifics of these processed datasets are presented in Table \ref{tab:table1}.

\noindent\emph{\textbf{Evaluation Metrics.}} We will use two widely utilized metrics, NDCG@$K$ and HR@$K$, to measure the model's capability, where $K$ is chosen as 5 and 10. For Top-$K$ recommendation tasks, we adopt a full-ranking evaluation strategy.

\noindent\emph{\textbf{Baseline.}} We select several advanced and related algorithms with different model structures as our baseline models. In traditional models, we use Caser~\cite{40}, BERT4Rec~\cite{43}, SASRec~\cite{35}, S3-Rec~\cite{44}, LightGCN~\cite{37}. For large model recommendation methods, taking into account both the size of baseline models and their language modeling capacities, we choose P5~\cite{1}, CID~\cite{Hua_2023}, IID~\cite{Hua_2023}, SID~\cite{Hua_2023} as our comparison model. Notably, there was a data leakage issue in the data preprocessing part of P5 that led to inflated metrics \cite{6}. Following the adjustments shown in \cite{6}, we modified the data preprocessing and conducted comparisons.
\begin{itemize}[leftmargin=*]
\item \textbf{Caser}: A CNN-based method, which captures high-order Markov chains for sequential recommendation by applying horizontal and vertical convolution operations.
\item \textbf{LightGCN}: This is a representative graph-based collaborative method that utilizes a simplified graph convolutional neural network to enhance user interest modeling.
\item \textbf{BERT4Rec}: A method based on a bidirectional self-attention model, with a cloze test target for the sequential recommendation.
\item \textbf{SASRec}: This is a representative sequential recommendation method that uses a self-attention network to encode sequential patterns for modeling user interests. It can be considered a collaborative method that takes sequence information into account.
\item \textbf{S3-Rec}: This method utilizes the intrinsic data correlation to derive self-supervised signals and pre-training to enhance the data representation of sequential recommendation.
\item \textbf{P5}: P5 utilizes the pre-training of LLM and transforms the recommendation task into custom natural language sentences using personalized prompts.
\item \textbf{CID}: CID premised on the assumption that items co-occurring more frequently are more similar, advocates for sharing a greater overlap of tokens during index construction. IDs are generated accordingly.
\item \textbf{IID}: IID entails creating an independent, additional tag for each item that requires learning, serving as the item's ID.
\item \textbf{SID}: SID leverages collaborative information via sequential indexing, assigning consecutive numeric IDs to items interacted with by users in sequence.

\end{itemize}

\begin{table}[t!]
\centering
  \caption{Statistics of the evaluation datasets.}
  \label{tab:table1}
  \begin{tabular}{cccc}
    \toprule
    Dataset&\textbf{Beauty}&\textbf{Toys}&\textbf{Yelp}\\
    \midrule
    \#Users & 22,363 &19,412 &30,431\\
    \#Items & 12,101 &11,924 &20,033\\
    \#Reviews & 198,502 &167,597 &316,354\\
    \#Sparsity(\%) & 0.0734& 0.0724&0.0519\\
  \bottomrule
\end{tabular}

\end{table}

\noindent\emph{\textbf{Implementation Details.}} We use the AdamW optimizer and adopt a warmup of 0.05, with a linear decay strategy for the learning rate. In terms of hyperparameter tuning, the learning rate is explored within the range of [1e-4, 1e-5], the temperature selection range is [0.05, 0.2], and the weight decay is set to 1e-3. For the embedding dimensions of all other recommendation methods, we choose 64. In the negative samples of contrastive learning, we refer to the Moco v1~\cite{46} paradigm, choosing a queue method for storing negative samples with a queue length of 512. For the LLM component, we select Flan-t5 \cite{45} because its excellent encoder can perform representation learning, and we also considered this factor when comparing other methods. We use the Nvidia A6000 GPU for training, with a batch size of 8, and adopt a gradient accumulation strategy, with an accumulation step of 8. Additionally, in the process of sampling second-order neighbor nodes, we ensure that the sampling of second-order neighbors for each first-order neighbor node is at a minimum of 8.
% \begin{table*}[t]
% \centering
% \caption{Results of the ablation studies over LLM with respect to GAL-Rec.}
% \label{table3}
% \setlength{\tabcolsep}{3pt}
% \begin{tabular}{clcccccccccccc}
% \toprule
% & \multirow{2}{*}{Methods} & \multicolumn{4}{c}{Sports} & \multicolumn{4}{c}{Beauty} & \multicolumn{4}{c}{Toys} \\
% \cmidrule(lr){3-6} \cmidrule(lr){7-10} \cmidrule(l){11-14}
% & & HR@5 & NDCG@5 & HR@10 & NDCG@10 & HR@5 & NDCG@5 & HR@10 & NDCG@10 & HR@5 & NDCG@5 & HR@10 & NDCG@10 \\
% \midrule
%  \footnotesize w/o  & \footnotesize GAL(SASRec) & 0.0266 & 0.0166 & 0.0429 & 0.0218 & 0.0453 & 0.0290 & 0.0705 & 0.0371 & 0.0453 & 0.0303 & 0.0661 & 0.0367 \\
%  \footnotesize w/o  & \footnotesize GAL(LGCN) & 0.0179 & 0.0113 & 0.0304 & 0.0152 & 0.0292 & 0.0189 & 0.0490 & 0.0252 & 0.0257 & 0.0169 & 0.0393 & 0.0211 \\
%  \footnotesize w/o  & \footnotesize GAL(Text) & 0.0220 & 0.0142 & 0.0345 & 0.0182 & 0.0456 & 0.0293 & 0.0705 & 0.0375 & 0.0354 & 0.0231 & 0.0554 & 0.0298 \\
%  \footnotesize w/  & \footnotesize GAL(SASRec) &\textbf{0.0282} & \textbf{0.0178}& \textbf{0.0452} &\textbf{0.0233} & 0.0486 & 0.0311 & 0.0759 & 0.0399 & \textbf{0.0491} & \textbf{0.0322} & \textbf{0.0712} & \textbf{0.0393} \\
%  \footnotesize w/  & \footnotesize GAL(LGCN) &  0.0203& 0.0128& 0.0328 & 0.0168& 0.0334 & 0.0214 & 0.0537 & 0.0279 & 0.0289 & 0.0190 & 0.0448 & 0.0241 \\
%  \footnotesize w/  & \footnotesize GAL(Text) &0.0251 & 0.0162& 0.0392 & 0.0211& \textbf{0.0492} & \textbf{0.0320} & \textbf{0.0768} & \textbf{0.0406} & 0.0388 & 0.0260 & 0.0587 & 0.0325 \\
% \bottomrule
% \end{tabular}
% \end{table*}
% \vspace{-1.em}
\subsection{Performance Comparison (RQ1)}
The experimental results of sequential recommendation are shown in Table \ref{table2}. The best results across all methods are denoted by bold numbers, and the second-best results by underlined numbers. For GAL-Rec's external embedding, we utilized LightGCN and SASRec as conventional recommendation models to capture collaborative information for one type of embedding, while also incorporating the item's textual information as another embedding type.
From the experimental results in the table, we can draw the following conclusions:
\begin{itemize}[leftmargin=*]
\item In comparison with conventional recommendation models, the GAL models utilizing their embeddings, such as LGCN and SASRec, have achieved substantial improvements across all datasets. As Zhao\cite{zhao2023embedding} highlighted, two fundamental aspects contribute to the effectiveness of recommendation models. Firstly, the embedding of items is pivotal; the caliber of item embeddings essentially determines the upper limit of the model's performance. Secondly, the modeling of users grounded in their historical interactions is critical(SASRec achieves this through attention mechanisms, whereas LightGCN leverages graph convolution operations.) The remarkable advancements credited to GAL can be primarily ascribed to its heightened capacity to model the intricate relationships within the user-item graph, thereby enhancing overall recommendation performance.
\item A significant positive correlation was observed between the performance of GAL-Rec, integrating diverse collaborative information, and the outcomes of various traditional models. This suggests that GAL-Rec's enhanced understanding of collaborative information is an extension of the intrinsic qualities of the collaborative information itself.
\item Notably on a specific dataset (Beauty), GAL-Rec featuring item text embedding (GAL-Rec (Text)) achieved the most superior performance. This implies that, in certain contexts, textual information can offer more effective insights.
\end{itemize}
\vspace{-1em}
\subsection{Ablation Study (RQ2)}

\linespread{1.25}
\begin{table*}[t]
\centering
\caption{Results of the ablation studies over LLM with respect to GAL-Rec.}
\label{table3}
\setlength{\tabcolsep}{3pt}
\begin{tabular}{clcccccccccccc}
\toprule
& \multirow{2}{*}{Methods} & \multicolumn{4}{c}{Yelp} & \multicolumn{4}{c}{Beauty} & \multicolumn{4}{c}{Toys} \\
\cmidrule(lr){3-6} \cmidrule(lr){7-10} \cmidrule(l){11-14}
& & HR@5 & NDCG@5 & HR@10 & NDCG@10 & HR@5 & NDCG@5 & HR@10 & NDCG@10 & HR@5 & NDCG@5 & HR@10 & NDCG@10 \\
\midrule
 \footnotesize w/o  & \footnotesize GAL(SASRec) & 0.0155 & 0.0078 & 0.0267 & 0.0113 & 0.0453 & 0.0290 & 0.0705 & 0.0371 & 0.0453 & 0.0303 & 0.0661 & 0.0367 \\
 \footnotesize w/o  & \footnotesize GAL(LGCN) & 0.0223 & 0.0152 & 0.0369 & 0.0.184 & 0.0292 & 0.0189 & 0.0490 & 0.0252 & 0.0257 & 0.0169 & 0.0393 & 0.0211 \\
 \footnotesize w/o  & \footnotesize GAL(Text) & 0.0309 & 0.0194 & 0.0425 & 0.0266 & 0.0456 & 0.0293 & 0.0705 & 0.0375 & 0.0354 & 0.0231 & 0.0554 & 0.0298 \\
 \footnotesize w/  & \footnotesize GAL(SASRec) &0.0216 &0.0154 &0.0329  &0.0198 &0.0486 & 0.0311 & 0.0759 & 0.0399 & \textbf{0.0491} & \textbf{0.0322} & \textbf{0.0712} & \textbf{0.0393} \\
 \footnotesize w/  & \footnotesize GAL(LGCN) & 0.0323 &0.0213 & 0.0457 & 0.0256 & 0.0334 &0.0214 & 0.0537&0.0279  &0.0289&0.0190 &0.0448  &0.0241  \\
 \footnotesize w/  & \footnotesize GAL(Text) &\textbf{0.0361} &\textbf{0.0258} &\textbf{0.0491}  &\textbf{0.0298}& \textbf{0.0492} & \textbf{0.0320} & \textbf{0.0768} & \textbf{0.0406} & 0.0388 & 0.0260 & 0.0587 & 0.0325 \\
\bottomrule
\end{tabular}
\end{table*}

In this section, to verify the contribution of graph-aware contrastive learning in the GAL-Rec framework, ablation studies were conducted, with the outcomes detailed in Table.

To substantiate the efficacy of the GAL framework, we conducted evaluations on all datasets and with various embeddings, comparing model performance under conditions with and without GAL integration. The latter scenario, representing the absence of graph-aware learning, entailed solely fine-tuning leveraging large language models. The outcomes presented in Table~\ref{table3} illustrate a 7\% to 15\% improvement across various metrics when employing GAL, affirming its significant impact on the effective fusion of LLMs and graph learning.

Delving further into the analysis of the GAL framework, Table~\ref{tab:lossablation} underscores the vital role each loss function plays. Moreover, insights from Table~\ref{tab:hopablation} indicate that an overly extensive receptive field could lead to less than optimal performance, emphasizing that a judicious choice of hop numbers for information aggregation is pivotal to achieving the best results. These findings collectively reinforce the notion that the careful calibration of the GAL framework, particularly regarding loss functions and graph traversal depth, is central to unlocking its full potential in enhancing recommendation system performance.
\begin{table}[t!]
\centering
  \caption{GAL (LGCN) ablation experiment under different loss(Toys).}
  \label{tab:lossablation}
  \begin{tabular}{ccccc}
    \toprule
    Hop&\textbf{H@5}&\textbf{N@5}&\textbf{H@10}&\textbf{N@10}\\
    \midrule
\ w/o L1 & 2.69 & 4.0 & 1.81 & 2.19 \\
\ w/o L2 & 2.77 & 4.25 & 1.80 & 2.26 \\
\ w/o L3 & 2.83 & 4.29 & 1.84 & 2.39 \\
\ ALL & \textbf{2.89} & \textbf{4.48} & \textbf{1.90} & \textbf{2.41} \\
  \bottomrule
\end{tabular}
\end{table}

\begin{table}[t!]
\centering
  \caption{GAL (LGCN) ablation experiment under different hop(Toys).}
  \label{tab:hopablation}
  \begin{tabular}{ccccc}
    \toprule
    Hop&\textbf{H@5}&\textbf{N@5}&\textbf{H@10}&\textbf{N@10}\\
    \midrule
\ 1-hop & 2.69 & 4.31 & 1.80 & 2.24 \\
\ 2-hop & 2.89 & \textbf{4.48} & \textbf{1.90} & \textbf{2.41} \\
\ 3-hop & \textbf{2.90} & 4.43 & 1.84 & 2.33 \\
  \bottomrule
\end{tabular}
\end{table}

\subsection{Distribution Uniformity Analysis (RQ3)}
One of the principal contributions of GAL-Rec is to enhance the understanding ability of LLM towards collaborative information. To better grasp how GAL-Rec enhances the encoding component's ability to represent user (item) data, we use KDE\cite{47} to visualize the embedding distribution of multi-hop information of users (items) in two-dimensional space. This visualization, demonstrated in Fig. \ref{fig:figure4}, highlights GAL-Rec's effectiveness in refining the embedding distribution. Furthermore, we also visualized the density of the representation of each multi-hop information on the unit hypersphere in terms of angles. Through this analysis, we observe:
\begin{figure*}[t]

\centering
\includegraphics[width=0.95\textwidth]{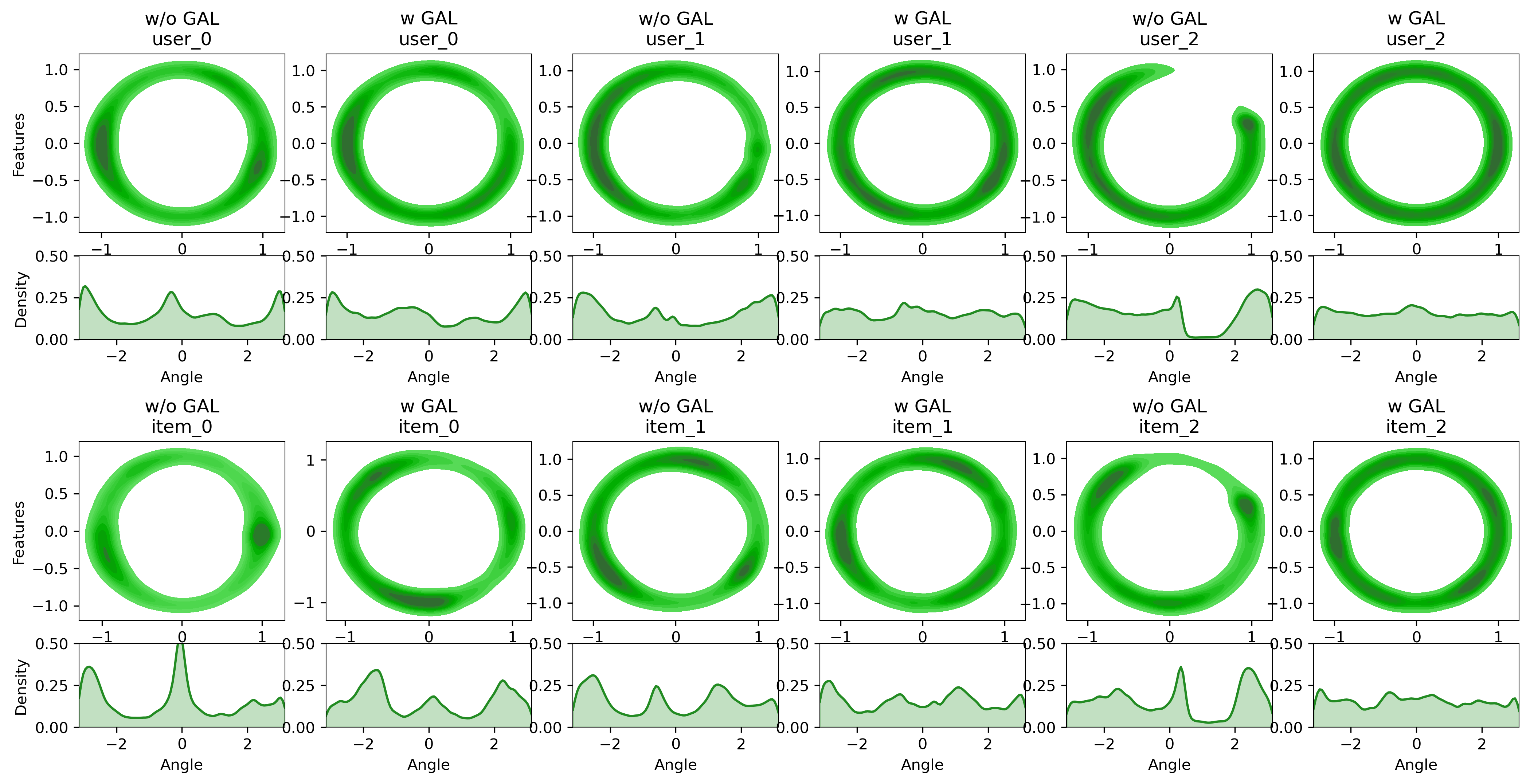}
\caption{Visualization: Feature distribution of multi-hop information representations of users(items) learn from Toys in two-dimensional space (The darker the color, the more users fall within that area).}
\label{fig:figure4}

\end{figure*}
\begin{itemize}[leftmargin=*]
\item In the absence of the GAL-Rec framework, multi-hop information exhibits uneven distribution. Simultaneously, with the expansion of the aggregation range, there is a notable collapse in distribution at the 2-hop aggregation information, where more user (item) representations tend to be similar. This phenomenon is akin to the over-smoothing issue observed in GNN, suggesting that LLM, when aggregating multi-hop collaborative signals, might face similar challenges. This insight provides a strategic direction for employing LLM in recommendation systems.
\item With GAL-Rec, the distribution of multi-hop information for users (item) is markedly more uniform. Particularly, GAL-Rec effectively addresses the collapse in representation at the 2-hop aggregation, underscoring the uniformity in signal representation after GAL-Rec processing. This indicates that GAL-Rec, by leveraging contrastive learning, can effectively model the interrelations among multi-hop collaborative signals of users (item), significantly boosting LLM’s capacity for processing collaborative information and enhancing its recommendation efficacy.
\end{itemize}
\subsection{Parameter Analyses (RQ4)}
In the GAL-Rec framework, we incorporate the concepts of contrastive learning temperature $\tau$, queue length $l$, and second-order neighbor count $n$. These hyperparameters' sensitivity will be evaluated through experimental investigation to enhance the applicability and development of GAL-Rec.

\begin{figure}[t!]
  \centering
  \includegraphics[width=\linewidth]{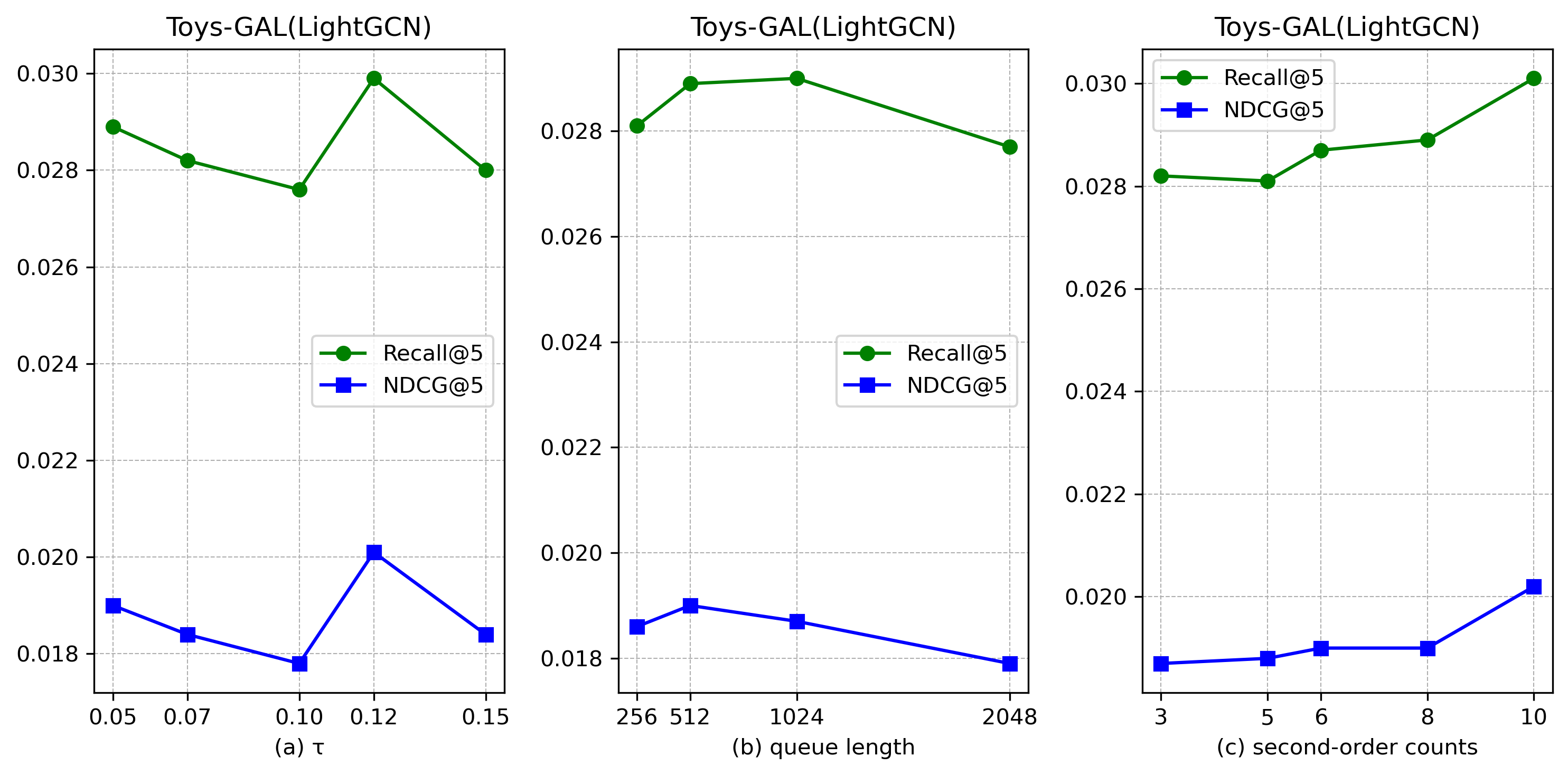}
  \caption{Effect of $ \tau$ and $length$ , $neighbors \ counts$ in GAL-Rec.}
  \label{fig:figure5}
% \vspace{-1.em}
\end{figure}

\emph{\textbf{ 5.4.1 Temperature.}} To examine the impact of the temperature parameter on GAL-Rec's performance, we adjusted its range. Fig. \ref{fig:figure5}(a) illustrates that the temperature parameter is pivotal for GAL's efficacy, with GAL-Rec's performance exhibiting significant fluctuations in response to variations in this parameter. The temperature parameter dictates the extent of focus on challenging negative samples by the contrastive loss mechanism. Choosing the GAL temperature parameter requires a balance: a higher parameter lessens emphasis on hard negatives, while a lower parameter focuses on similar negatives, aiming for a more uniform representation space.

\emph{\textbf{5.4.2 Queue Length.}} A series of experiments were conducted to assess the queue length's impact on GAL. Initial results, as depicted in Fig. \ref{fig:figure5}(b), indicated that increasing the queue length enhances GAL's performance; however, it does not significantly improve the model's performance. Moreover, increasing the queue length beyond a certain point did not yield additional performance gains and actually diminished the model's efficacy, which contrasts with the findings associated with Moco v1. However, considering the cancellation of the memory bank module in contrastive learning based on transformers in Moco v3~\cite{49}, we believe that the queue length has lost its previously significant effect on model performance in transformer structures.

\emph{\textbf{5.4.3 Second-Order Neighbors Count.}} To investigate the impact of the number of second-order neighbors on the performance of GAL, we varied the range of second-order neighbor counts. As illustrated in Fig. \ref{fig:figure5}(c), the performance of GAL exhibits a steady upward trend with the increase in the number of second-order neighbors. This trend indicates that GAL benefits from the continuously expanding receptive field, underscoring the significant role of aggregating information across multiple hops within the GAL architecture and demonstrating the efficacy of our contrastive learning approach.

\section{CONCLUSIONS}
In this paper, we introduced GAL-Rec, a framework designed to enhance the recommendation capabilities of Large Language Models (LLMs). GAL-Rec leverages the principles of Graph Neural Networks (GNNs) to aggregate multi-hop information and employs contrastive learning to connect multi-hop user information with multi-hop item information. This approach enriches the understanding of collaborative semantics between users and items. Our experimental results demonstrated that GAL-Rec significantly improves performance. Additionally, to address the issue of inadequate negative samples in contrastive learning for LLMs, we propose a dynamic queue storage methodology based on Moco.

Looking ahead, we will explore the application of our method to Decoder-only structured LLMs. We also aim to extend our approach to a wider range of recommendation tasks, such as multimodal recommendation, click-through rate (CTR) prediction, and rating prediction tasks.

Overall, we have contributed a self-supervised learning paradigm for LLMs in recommendation systems. We believe that this research opens new avenues for the application of LLMs within recommendation systems.

 % argument is your BibTeX string definitions and bibliography database(s)
%\bibliography{IEEEabrv,../bib/paper}
%

\bibliographystyle{IEEEtran}
%\bibliography{IEEEabrv,sample-base.bib}
\bibliography{bare_jrnl_new_sample4.bib}

\newpage

\end{document}